\documentclass[aps, pra, a4paper, amsfonts, amssymb, amsmath, reprint, showkeys, twoside]{revtex4-2}
\usepackage{dcolumn}
\usepackage{ragged2e}
\usepackage{mathtools}
\usepackage{physics}
\usepackage{xcolor}
\usepackage{graphicx}
\usepackage[left=23mm,right=13mm,top=35mm,columnsep=15pt]{geometry} 
\usepackage[T1]{fontenc}
\usepackage[utf8]{inputenc}
\usepackage{csquotes}
\usepackage[english]{babel}
\usepackage{caption}
\usepackage[colorinlistoftodos, color=green!40]{todonotes}
    
\makeatletter
\def\maketitle{
\@author@finish
\title@column\titleblock@produce
\suppressfloats[t]}
\makeatother

\begin{document}

\title{Raman Signal Enhancement in Graphene via a Micro-Ring Resonator}

\author{A. Sharma$^{1}$, Y. Li$^{2}$, M. K. Prasad$^{3}$, W. L. Ho$^{4}$, S. T. Chu$^{4}$, I. V. Borzenets$^{1}$}
    \email[Correspondence email address: ]{borzenets@tamu.edu}
    \affiliation{$^{1}$Texas A\& M University, College Station, Texas, USA\\ $^{2}$Zhejiang Sci-Tech University, Hangzhou, China \\  $^{3}$Newcastle University, Newcastle upon Tyne, England
 \\ $^{4}$City University of Hong Kong, Kowloon Tong, Kowloon, Hong Kong SAR }

\begin{abstract}
\noindent Micro-ring resonators (MRRs) ``trap'' incoming light, and therefore, have been shown to achieve extremely high local intensities of light. Thus, they can be used to facilitate highly non-linear optical signals. By embedding materials that host non-linear optical processes inside the MRR, we expect to observe an enhancement in the strength of the non-linear optical signal. This concept is demonstrated here by extracting the Raman signature of graphene that is placed inside a MRR. A highly doped silica MRR which features an optical bus waveguide coupled to a loop (ring) tuned to near-infrared wavelengths is used. Raman signal with an excitation wavelength of $522$~nm via third harmonic generation inside the MRR is observed. Higher order Raman signal of the embedded graphene at the $1597.6$~nm excitation wavelength is also observed. This work demonstrates the feasibility of the MRR as a non-linear signal enhancer using novel MRR device setups.
\end{abstract}

\maketitle

\section{Introduction} 
\noindent Optical micro-ring resonator (MRR) devices play an important role in the world of integrated optics owing to its various capabilities such as field enhancement, optical filtering, modulation and more~\cite{alsing2017,Geuzebroek,Li2008}. An add-drop MRR consists of a ring resonator placed in-between two waveguides. At resonance, the signal coupled from the input waveguide to the MRR builds up in the MRR cavity and couples to the output waveguide. At off-resonance, the input signal will simply pass by the MRR without any intensity build-up in the cavity. The build-up of the evanescent field around the ring enables sensing, trapping, and illumination techniques. For example, sensing is facilitated by the resonance shift when the field interacts with an entering sample~\cite{Kwon2008}, pico-Newton forces generated by the fields can manipulate or trap small-sized particles~\cite{Ho2020}, and the field can also initiate photo-chemical reactions and photo-physical effects through illumination. Constructive interference within the ring allows retention of high local intensities compared to a straight bus optical waveguide~\cite{rabus2020}. The loaded Q-factors of currently available silicon-based ring resonators range from $10^5$ to $10^6$, making them an ideal platform for researching their use in  enhancement of non-linear signals~\cite{Sinclair2020}. As a result of high Q-factor which manifests very narrow resonance peaks in spectral reading, detection of change in refractive index and the peak have now become the standard for MRR-based sensing~\cite{Bryan2023}. In addition, due to advances in fabrication processes for silicon-based MRRs, they have been involved in various chemical and biological sensing with increasing sophistication~\cite{Moss2013,van2016,Bawankar2021}.\\
\indent Non-linear, high-order optical phenomena generally require high intensities to be seen. These phenomena are widely exploited to characterize a material's structural properties~\cite{Johansson2018,Belkin2005}. A particular example is Raman spectroscopy. Raman scattering in a material provides detailed compositional and structural information~\cite{White2006}. Raman signal intensities are of several order magnitude (>$10^6$) smaller than that of excitation laser source, hence enhancement of Raman signal is a good benchmark to study with MRRs~\cite{Larkin2011}. MRR's capacity to host high local intensities enables enhancement of Raman signal as Raman scales non-linearly with excitation intensity~\cite{Nikitin2013}.\\ 
\indent Presented here is an application of MRR's enhancement capacity, using two measurement schemes to conduct Raman spectroscopy of graphene. First measurement scheme under setup 1 [Figure 1 (a)] makes use of a MRR device where its waveguides are closer to surface and generates $5~\mu$W green light at $522$~nm via third harmonic generation (THG). With graphene deposited on top of this device, Raman signal of comparable intensity to conventional Raman spectroscopy from $20$~mW ~$532$~nm laser is observed. Second measurement scheme under setup 2 [Figure 1 (b)] makes use of a near-infrared (NIR) laser as excitation source. Here, we embed graphene directly into the MRR waveguide facilitated by trenching into the ring from surface [Figure 2 (e)], and observe Raman signal on spectra taken around $1597.6$~nm from the output waveguide. In the second setup, we do not see a signal without ring resonator.
\begin{figure*}
\centering
\includegraphics[width=\textwidth]{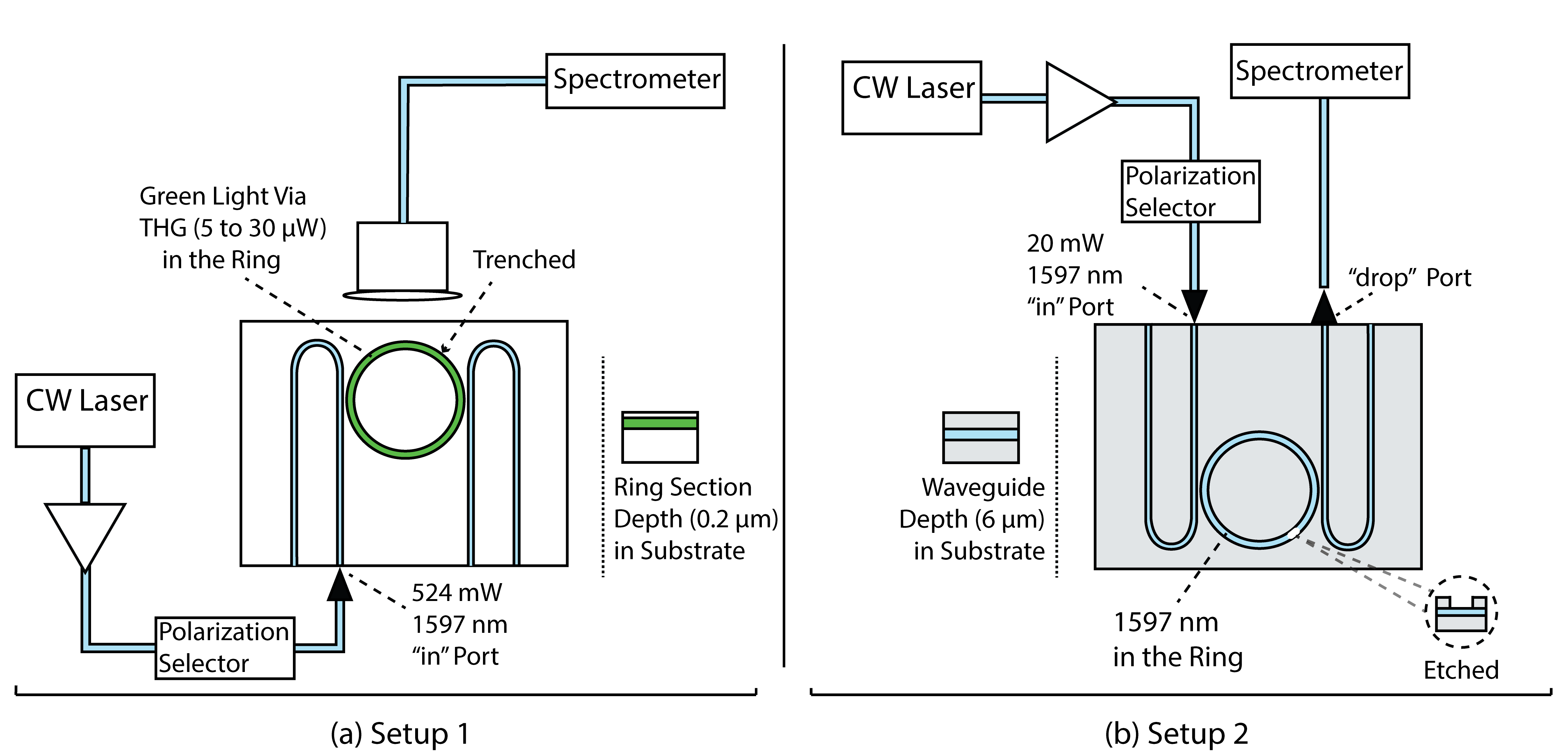};
    \caption{\justifying
    (a) Here setup 1 in first measurement scheme is shown. A C-band tunable laser source, passed through an optical amplifier then polarization selector, is connected to ``in'' port. The bus is evanescently coupled to the ring where green emission is generated via THG due to high intensities sustained in the ring. This $522$~nm THG emission produced at $1600.35$~nm input laser, acts as the excitation for samples deposited on top of the ring. (b) Here setup 2 in second measurement scheme is shown. A C-band tunable laser source is connected to the  ``in'' port of the MRR along with other elements as described for setup 1, although the laser source can directly be connected to the input on this measurement. Measurements can be taken from either ``add'', ``drop'' or ``through'' port of the straight bus by coupling a spectrometer probe directly to the ports using optical fibre. Characteristic Raman signals of the sample are observed in both cases as shown in Figure 3).}
\end{figure*}
\section{Experimental Setup} 
\noindent The general schematic of devices used in this experiment is a four-port add-drop MRR which consists of two bus waveguide sides coupled to a $135~\mu$m radius ring-resonator having a gap separation of $0.8~\mu$m. The silica clad waveguides have core dimensions of $2~\mu$m $\cross$ $1~\mu$m with core refractive index of $1.70$, and they are fabricated using CMOS-compatible processes~\cite{6058715}. The device was used in first demonstration of visible emission via THG and has been extensively characterized previously~\cite{li2020}. Measurements were taken on two variations of this device. Note that the MRR devices were connected to conventional Raman spectrometer system (WITec RAMAN alpha 300R) equipped with $532$~nm laser and spectrometer (UHTS 300 VIS-NIR) suitable for excitation in the range $532$~nm to $830$~nm. The system was modified to accept signals from MRR by bypassing the system's laser for setup 1, and by coupling output waveguides via optical fibres for  setup 2.\\
\indent The waveguides in device 1 [Figure 1 (a)] are covered with $6~\mu$m thick oxide as upper cladding, effectively putting the waveguides at $6~\mu$m depth from the surface. The oxide upper-cladding on the ring section is trenched such that the ring-waveguide is approximately $0.2~\mu$m from the surface. This structure makes the ring section easily accessible for sample placement either via deposition or in a solution form. We rely on THG from this device in order to achieve excitation wavelengths compatible with our existing commercial Raman setup. The waveguides being closer to the surface enables measurement of upward emission from the ring via a collimator lens connected to a spectrometer. Visible emission from the device can be used as the excitation source to study Raman signatures around this wavelength and demonstrate enhancement of non-linear signals within the ring-waveguide. A tunable continuous-wave (CW) laser source (Amonics, ATL-C-16-B-FA) was adopted as the pump source, followed by an erbium-doped fiber amplifier (EDFA, Amonics, AEDFA-30-R-FA) to amplify the NIR pump power. The pump wavelength was tuned to be $1600.35$ nm, with in-waveguide power measured to be $524.3$ mW, after deducting the fiber-waveguide coupling loss. Because the resonance point of the device can shift due to thermal activity, the device is also thermally stabilized using a Proportional–Integral–Derivative (PID) controlled Peltier cooler. Signal is detected via a microscope objective placed $5$ mm above the ring resonator to capture photons resulting from THG emission at $522$ nm as well as signals that result from interactions between emitted wave and the sample. Note that THG emission is not exactly one-third of the input wavelength due to thermally induced resonance shifts~\cite{li2020}.  We measure spectra at several spots on the ring [Figure 2 (b, c, d)]. Because of the nature of placement of scope, one can obtain positional resolution in this setup; that is, by translating microscope objective's position, one can gather spectra at different sections of the ring's surface. Using green light generated via THG as excitation source enables us to compare the results from the MRR with that of conventional Raman, excited by $532$~nm laser.\\
\begin{figure}\centering
\includegraphics[width=0.40\textwidth]{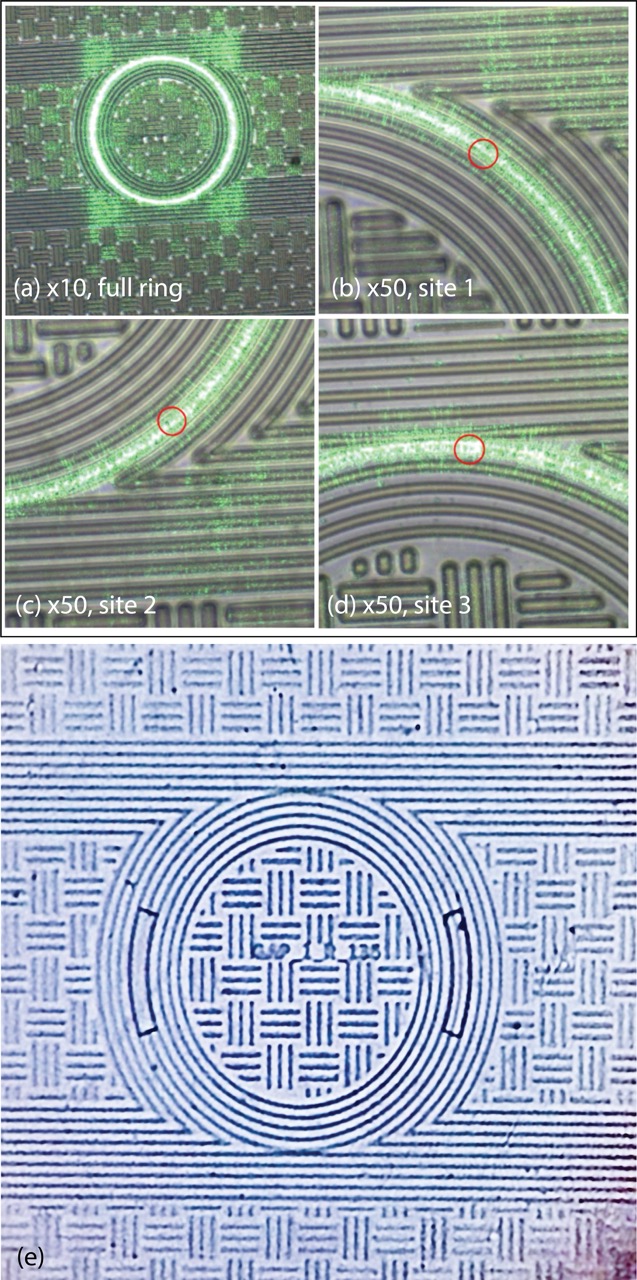}\\

    \caption{\justifying Image of a ring resonator device at various magnification while connected to a laser source. (a) Visible Green light due to $3\times frequency$ emission produced via THG is seen. (b,c,d) Red circles point to various sites, that were probed with the spectrometer. Green emission's power ranged from $1~\mu$W to $5~\mu$W during the time of imaging. (e) Image shows the two trenched area on the left and right of the ring section of a MRR device. Samples are deposited in these trenches.}
\end{figure} 
\indent On the second setup using device 2 [Figure 1 (b)], the waveguides are at standard depth under $6~\mu$m oxide cladding. To access the ring-waveguide, a hole is etched onto the surface. This opening acts as the sample deposition area, which in our case is a solution consisting of graphene flakes [see Supplementary for transfering process]. The measurement can be taken from either of the ``through'', ``add'' or ``drop'' port by directly coupling to the spectrometer. We label the port connected to source as ``in'' and the port connected to spectrometer as ``drop'' [Figure 1 (b)]. Here, we obtain aggregate Raman signal but not positional information. A tunable CW laser source (Amonics, ATL-C-16-B-FA) laser source held at $1597.6$~nm is coupled to the ``in'' port of the straight bus. The input power was set to $20$~mW. With no up-conversion inside the ring, the excitation wave for the sample is $1597.6$~nm. The detector is directly coupled to the ``drop'' port of the straight bus. Our available Raman spectrometer is only specified to operate in $532$~nm to $830$~nm excitation range, therefore, we must look at anti-Stokes scattering at large Raman shifts in order to detect a signal using this measurement scheme [see Supplementary for measurement limitations].\\
\indent Raman signatures from graphene have been extensively studied in past~\cite{ferrante2018,ferrari2013,Ferrari2006}. A single layer graphene shows Raman shifts at various wavenumbers. 2D peak, around $ 2580$ cm$^{-1}$, manifests as overtone to the D peak. The D peak is from transverse optical phonons present around the corner of Brillouin zone. 2D peak is from a process where two phonons with opposite wave vectors are involved, therefore it is a second-order non-linear process. D peak requires presence of defects, but 2D peak is always present~\cite{ferrari2013}. The intensity of 2D peak is also independent of excitation frequency, however the peak's position is dependent on excitation frequency. There are also multiple other higher order peaks in graphene as a result of multi-phonon scattering processes involving 3 or more phonons~\cite{Rao2011}.\\
\section{results} 
\begin{figure*}
 \centering
   \includegraphics[width=\linewidth]{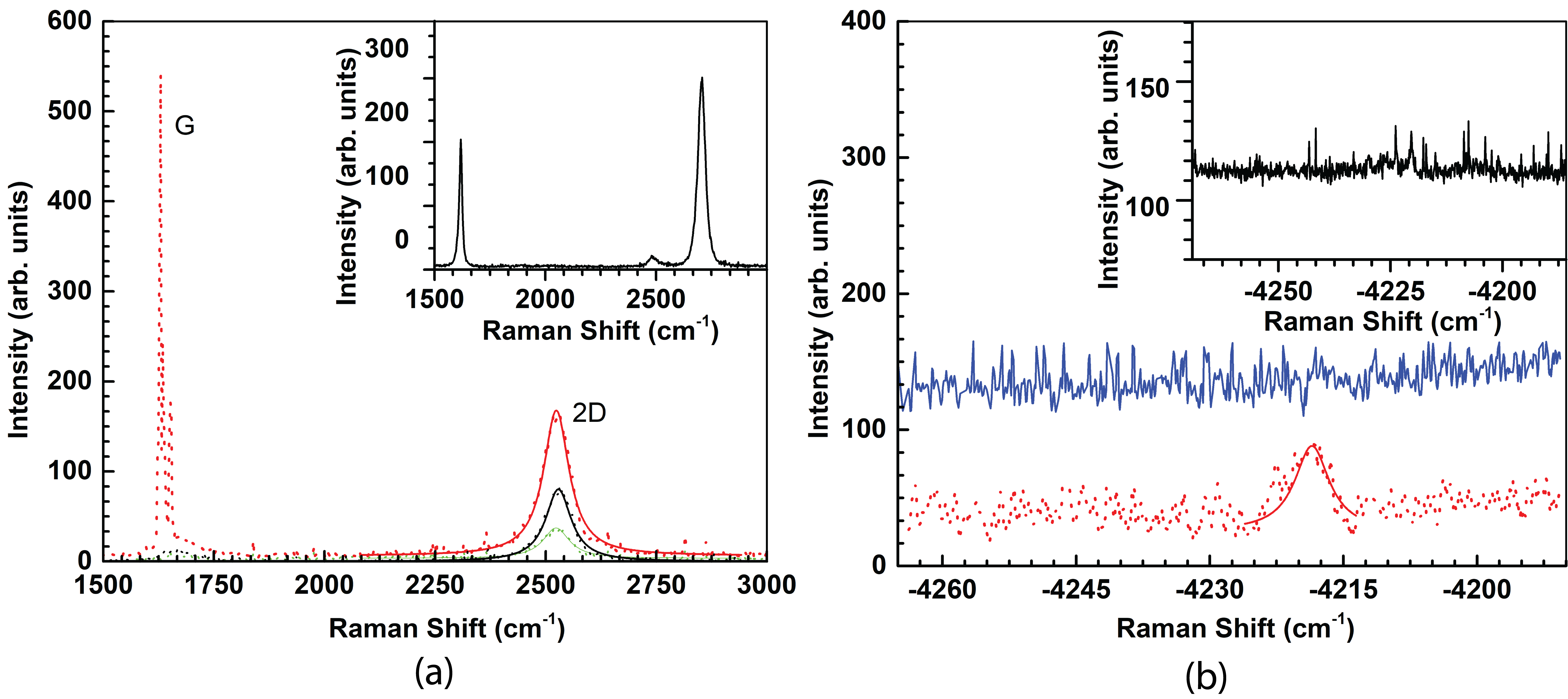}
\caption{\justifying Raman spectra extracted from different locations on MRR. (a) Spectra from site 1,2 and 3 are represented by scatter plots of different colors. Solid curves represent Lorentzian fitted to 2D Raman shift at these sites. Spectra show peaks around $2530$ cm$^{-1}$. Input laser wavelength of $1600.35$~nm is converted via third harmonic generation to $522$~nm in MRR, which is used as excitation source. The sample in use is graphene which is transferred onto the first device's ring via standard transferring techniques. The peaks around $2530$ cm$^{-1}$ correspond to graphene 2D peaks. Inset shows Raman spectra of graphene on silicon substrate using $20$~mW laser power. While the effective excitation power on the device is in between $5~\mu$W and $30~\mu$W. The Raman intensities are comparable even though the excitation power on the device is lower by a factor of $4*10^{3}$. This shows enhancement of the signal due to the resonator. (b) Here, the laser source is held at $1597.6$~nm on the second device, and the spectrum is scanned around same excitation wavelength. Measurements (red dot) fitted to a Lorentzian (red curve) show spectral peak around $-4220$ cm$^{-1}$, associated with higher order phonon processes. Blue spectra, shown with an offset, represents measurement taken from MRR without any graphene; it gives us background information which shows no features around $-4220$ cm$^{-1}$, thereby enabling us to deduce that the peak on the red curve is due to presence of graphene. In addition, the inset shows spectra taken from a device which only contains ``spiral'' waveguide with graphene deposited on it; no ring resonator is present. Here too we do not see any features because there is no enhancement without the resonator. }
    \label{fig:Raman signals from 3 different sites on MRR}
\end{figure*}
\noindent From the measurements, we want to demonstrate, by extracting Raman signatures, that MRR is capable of enhancing non-linear signals. As illustrated in figure 1, spectral measurements from two different devices were taken. First device makes use of THG green emission as excitation source to induce Raman scattering in the sample, while the second device uses the NIR laser as excitation source. \\
\indent On measurement setup 1, the MRR device 1, as shown in Figure 1 (a) is tuned to generate green light as the available Raman spectrometer was operable in the range $532$~nm to $830$~nm excitation wavelengths. Three sites were probed on the ring as labeled in Figure 2 (b,c,d). One can observe clear Raman signals of graphene from the spectra [Figure 3 (a)]. In all cases we observe 2D peaks of graphene with intensity variation, and G peak with noticeable intensity at site 1. Here, 2D peaks are identified consistently across all data sets. In particular, locations of 2D peaks for the fitted Lorentzians are $2528.7$ cm$^{-1}$ on site 3, $2523.5$ cm$^{-1}$ and $2524$ cm$^{-1}$ on site 1, $2528$ cm$^{-1}$ on site 2 for excitation wave $522$~nm in the ring as a result of THG up-conversion of $1600$~nm input laser. The power of the upward emitted green light for the whole ring was measured to be between $5~\mu$W and $30~\mu$W. To benchmark this setup, we also study graphene grown on copper by chemical vapor deposition (CVD) that is transferred onto a silicon substrate via lifting and transferring technique~\cite{Suk2011}. Conventional Raman spectroscopy was performed on this graphene on Si sample with $20~m$W laser power using the WITec Raman system, see inset of Figure 3 (a). The measured Raman intensities are comparable. This is likely because even though total power of green emission was $5~\mu$W, local intensity on the site of sample is much higher in the MRR device.\\
\indent On measurement setup 2, the MRR device 2 is used [Figure 1 (b)]. With $1597.6$~nm excitation, the spectra from the ``drop'' port was measured. We chose to directly couple spectrometer to the drop port for following reasons: this enables us to get aggregate signal without the use of a microscope; existing lenses on WITec Raman system are not compatible at NIR wavelengths.  The results show evidence of detection of a peak as a result of  higher order multi-phonon process which are weaker in intensity. Specifically, we present detection of G+2D peak at around $-4220$ cm$^{-1}$ shift~\cite{Rao2011}. We chose anti-Stokes G+2D signal as this is the only one in the WITec system spectrometer's range; anti-Stokes shifts, having shorter wavelengths fall in the range of spectrometer. The data presented here is after minimal processing: single point noise-peaks were removed. The peak at $-4220$ cm$^{-1}$ is present on MRR device with graphene as shown in red Figure 3 (b), but is not seen in empty MRR device data as shown in blue. In addition, we prepared devices where the waveguides are spirals and do not involve a ring resonator. Graphene was deposited onto this spiral-based device [Supplementary Figure 1 (b)]. Measurements from output of the spiral waveguide directly coupled to spectrometer were taken with laser input held at $1597.6$~nm into waveguide. We do not observe any peak on the spectra taken from spiral-based device, see inset of Figure 3 (b). We noticed parasitic Raman signal from the optical fiber as well as from the erbium-based amplifier attached to the input [see Supplementary for noise and artifacts discussion]. Non-identifiable 2D and G peaks are attributed to the combination of lower excitation frequency in use, parasitic signals from fibre as well as the amplifier, and spectrometer being out of range: parasitic signals from the erbium doped amplifier overwhelms the spectral points with strong activity where the usual G, 2D Raman signatures are present.\\
\section{Conclusion}
\noindent Here we utilized MRRs to perform Raman spectroscopy on graphene, utilizing two different measurement schemes on two variations of the device, to demonstrate non-linear signal enhancing capacity of MRR. Raman shifts were targeted to be studied since they are non-linear, higher order signals. Both schemes show signatures of peaks due to Raman emission. First measurement scheme with setup 1 enables positional data around the ring, whereas the second measurement scheme with setup 2 provides aggregate Raman signal of the sample. The effective power on the ring on setup 1 was seen to be between $5~\mu$W and $30~\mu$W, although the local intensity at sample location is likely higher. These measurements show that Raman signal can be extracted using very low total intensities, which paves a way to perform spectroscopic analysis on various organic materials without causing sample burn. This research therefore demonstrates feasibility of using a MRR for signal enhancement and spectroscopic analysis.
\section{Acknowledgement}
\noindent S. T. Chu acknowledges support from the City University of Hong Kong (APRC Grant no. 9610395). Y. Li acknowledges the National Natural Science Foundation of China (No. 62105291). This work was supported by the Strategic Transformative Research Program (STRP) at Texas A\&M University (I. V. B. and A. S.).

\clearpage
\newpage
 
\setcounter{table}{0}
        \renewcommand{\thetable}{S\arabic{table}}%
        \setcounter{figure}{0}
        \renewcommand{\thefigure}{S\arabic{figure}}%
        \setcounter{section}{0}

\title{Supplementary to Raman Signal Enhancement in Graphene via a Micro-Ring Resonator}
\maketitle
\setcounter{page}{1}
\onecolumngrid
\section{MRR and Spiral Devices}
Raman spectroscopy of graphene was performed using
two variations of device, utilizing two measurement
schemes. Optical image of a standard micro-ring resonator (MRR) device is
shown here in Figure 1. The ring radius is 135 $\mu$m and the silica clad waveguides have dimensions of $2~\mu$m$\times$1$~\mu$m. The upper cladding is 6 $\mu$m thick, and the refractive index of the core is at 1.70. Here we further elaborate on the sample deposition process, measurement limitations, and noise sources. Despite measurement limitations, we present observation of detection of a peak centered at $-2610$ cm$^{-1}$
\begin{figure}[ht]
    \centering
\includegraphics[width= 0.8\linewidth]{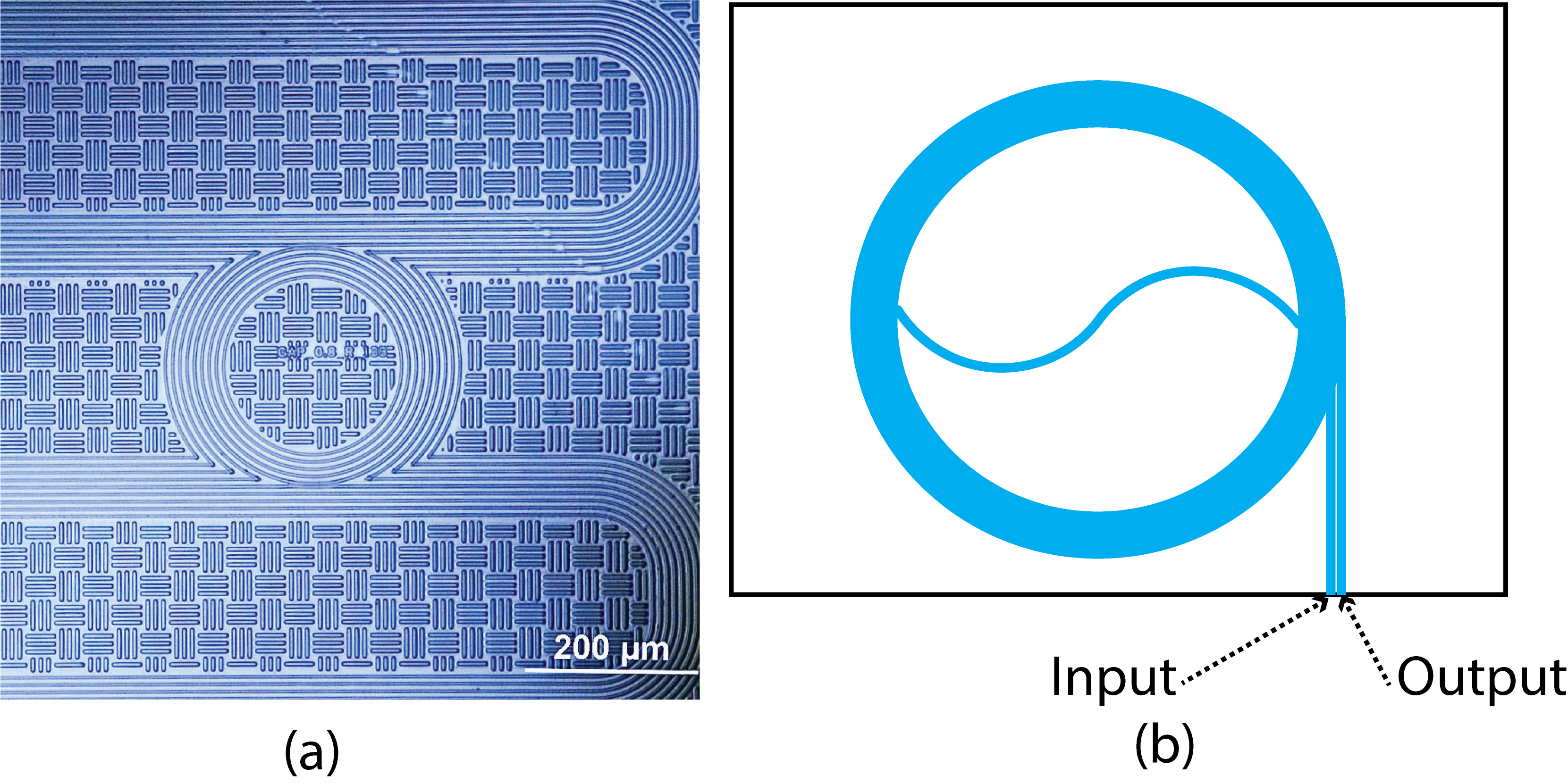}
    \caption{ (a) Microscope picture of the MRR device used in setup 1 and setup 2 described in the main text. As seen here on top section of the image, a waveguide of dimensions $2~\mu$m$\times$1$~\mu$m couples to ring waveguide of radius $135~\mu$m. A second waveguide of same dimensions also couples to the ring, as seen here on the bottom section of the image. These straight waveguides act as input and output ports of the device. (b) Schematic of a device with spiral waveguide (blue) used in benchmarking measurements described in the main text results and supplementary. The waveguide starting from the "input" port spirals inward and makes an "S" shaped path at the center, after which it  spirals outward along a similar path and ends with an "output" port. }
\end{figure}
\section{Graphene Transfer Process}
 Chemical vapor deposition (CVD) grown graphene as a sample was deposited onto the devices in two different forms; straight deposition, or as a solution. In case of device 1 the sample can be deposited using both standard wet-transfer process~\cite{Suk2011sup,Ullah2021sup}, and as a solution, since the ring-waveguide with excitation signal is accessible to the deposited sample on the surface which is only $0.2$ $\mu$m from ring. The wet-transfer process is described as follows: Commercially available CVD graphene on copper foil was obtained. $1$ cm$^{2}$ strip of the foil is cut and spin coated with polymethyl-methacrylate (PMMA) resist (495K PMMA A4, Microchem) at 3000 RPM for 60 seconds. CVD graphene is present on both sides of the foil, hence to etch way the unnecessary graphene, the back of the foil was exposed to oxygen plasma for 10 to 15 seconds. After leaving the coated foil to dry on hot plate for 20 minutes at 105 Celsius, the foil was floated on top of copper etchant for 2 to 3 hours. After the copper layer was completely etched, the graphene with PMMA was cleaned with deionized water and deposited on top of the device aiming for the ring section. To develop a graphene solution, graphene flakes were mixed in acetone and the resulting solution can be dropped into the etched hole or onto the ring section for measurements.
\section{Determination of THG Intensity and Frequency}
For benchmarking purposes, spectra from device 1 were collected using the collimator F671SMA from Thorlabs which is sent via optical fiber cable to the optical spectrometer USB2000+ from Ocean Optics with 0.38 nm step size for determining both the intensity and frequency of the green light signal generated via third-harmonic generation (THG). The calibration was done using a reference silicon detector DET100A from Thorlabs, between $400$ nm and $1100$ nm. During the experiment, a continuous wave laser, Amonics ATL-C-16-B-FA, held at 1600.35 nm  wavelength was used as excitation source in conjunction with AEDFA-30-R-FA EDFA. From device 1 utilizing measurement scheme 1, the raw spectra data show consistent 2D peaks, whereas G peak is only seen at site 1~\cite{Ferrari2006sup,ferrari2013sup}. Figure 3 (a) in the main text thoroughly discusses this.\\
Additionally, we studied Raman signals from graphene on silicon (Si) using $20$ mW $532$ nm laser. See main text Figure 3 (a) inset for comparisons between proposed measurement setup 1 and the standard Raman spectroscopy of graphene on Si. We also performed measurements on other variations of devices, with and without sample deposition to benchmark our experimental setup, most of which are covered throughout the text. The variations were: device with a ring but no graphene, device with spiral waveguides no ring-resonator and no graphene deposition, device with spiral waveguides with graphene deposition.  Schematic of a device with a spiral waveguide is shown here in Figure S1 (b).
\section{Measurement Limitations}
 Measurements were taken using a Raman spectrometer system (WITec RAMAN ALPHA 300R) equipped with $532$~nm laser and spectrometer (UHTS 300 VIS-NIR) suitable for excitation in the range $532$~nm to $830$~nm. The system was modified to accept signals from the MRR device by bypassing the system's laser for setup 1, and by coupling the device's output waveguides via optical fibres to the spectrometer for setup 2. The spectrometer is only specified to be suitable for $830$ nm laser excitation which means the sensing capability of the spectrometer sensor for excitation wavelengths greater than $830$ nm is inefficient, hence measured signals using setup 2 at $1550$ nm excitation are highly suppressed. Furthermore, we cannot collect upward emissions from the ring at $1550$ nm range due to microscope objectives in the Raman spectrometer system being only specified to work around $830$ nm at maximum. The specification of the objective states that the photon transmission for wavelengths greater than 1120 nm is suppressed by 50\%, which extrapolates to less than 30\% effective transmission above $1270$ nm wavelength photons. 
\begin{center}
\begin{tabular}{ |c|c|c| } 
 \hline
 Raman Shifts at $1550$ nm Excitation & Output Wavelength & Objective Transmittance \\ 
 \hline
 $-1600$ cm$^{-1}$ & $1273$ nm & $< 30\%$ (Extrapolated) \\ 
 $ -2610$ cm$^{-1}$ & $1129$ nm & $< 50\%$ \\ 
 $ -2700$ cm$^{-1}$ & $1117$ nm & $\approx 50\%$ \\ 
 $ -4220$ cm$^{-1}$ & $955$ nm & $\approx 75\%$ \\ 
 \hline
 \multicolumn{2}{|c|}{Max Specified Spectrometer Excitation}  & $830$ nm\\
 \hline
\end{tabular}
\label{Table S1: Objective transmittance}
\end{center}
\section{Excitation Source Artifacts}
\begin{figure}[h]
    \centering
    \includegraphics[width=\linewidth]{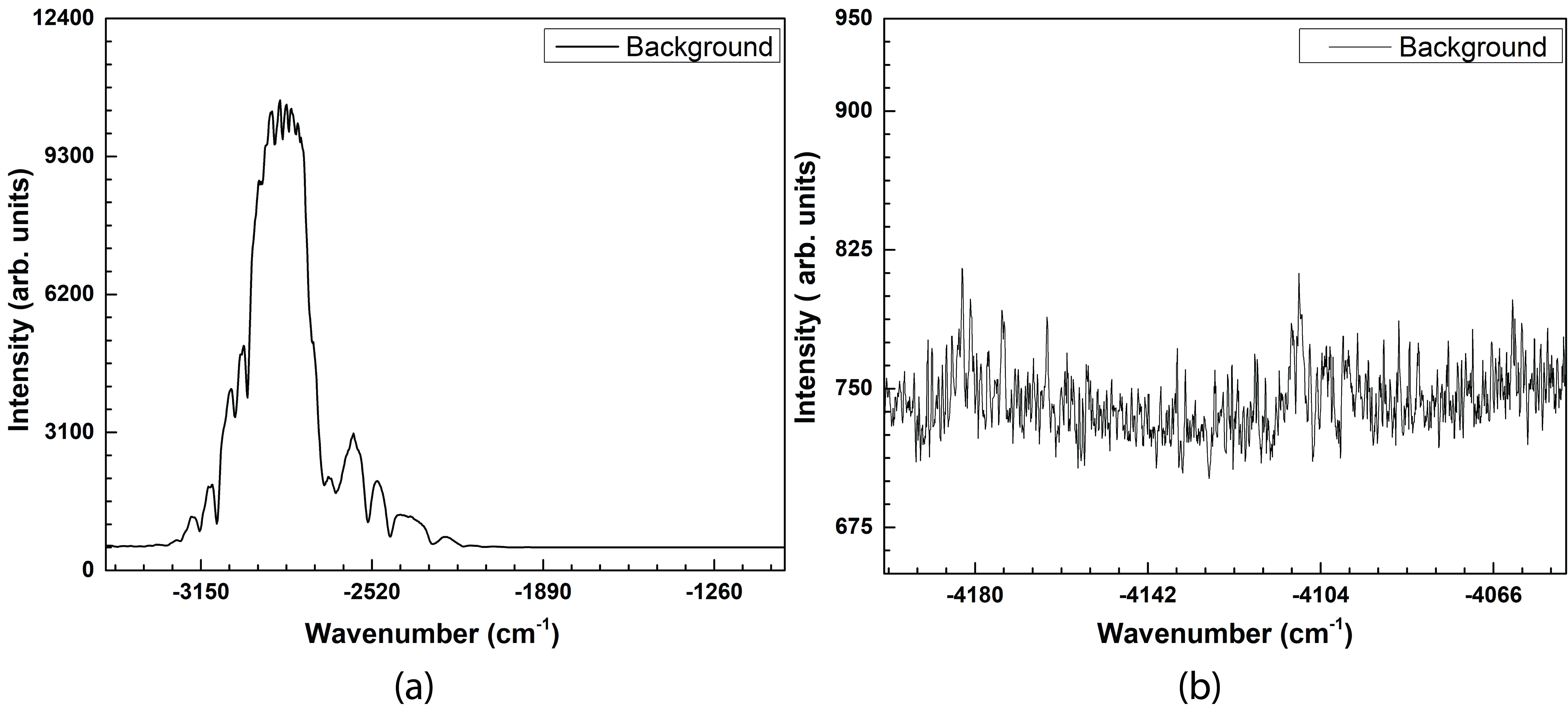}
    \caption{ (a) Spectra taken from the ``drop'' port of the MRR device without any sample at $1600$ nm excitation. The laser is passed through an amplifier and fed into the device via optical cables. The source of artifacts in the spectra that ranges from $-2500$ cm$^{-1}$ to $-3200$ cm$^{-1}$ is attributed to erbium-doped fibre amplifier(EDFA). (b) Spectra taken from the ``drop'' port of the MRR device without any sample at 1549 nm excitation. Observed artifacts at $-4100$ cm$^{-1}$ and $-4180$ cm$^{-1}$ in this measurement are also attributed to EDFA.}
    \label{fig:Noise Profiles}
\end{figure}
Measurements are made with the MRR device connected to a laser source which has been passed through an amplifier (Setup 2) and also a polarization selector (Setup 1). The amplifier is an essential device which provides high input power for the MRR to function when using proposed measurement schemes. Here we report on the presence of artifacts due to the amplifier in spectra collected at certain wavelengths, particularly, the artifacts that directly prevent us from recording the anti-Stokes 2D graphene peak located around $-2600$ cm $^{-1}$ region but does not interfere with the signal at $-4220$ cm$^{-1}$. Figure S2 (a) shows persistent background noise from source that spreads from $-2500$ cm$^{-1}$ to $-3200$ cm$^{-1}$. Figure S2 (b) shows presence of background noise from source at $-4100$ cm$^{-1}$ and $-4180$ cm$^{-1}$. That these are indeed source artifacts is verified by connecting the laser and amplifier directly to the Raman spectrometer via a short optical cable. These artifacts still persist when probing the source (laser and amplifier) using  AQ6370D optical spectrum analyzer, but are not present when only the laser was used. 

\section{Graphene 2D Peak with 1600 nm Excitation}

\begin{figure}[ht]
    \centering
    \includegraphics[width=0.5\linewidth]{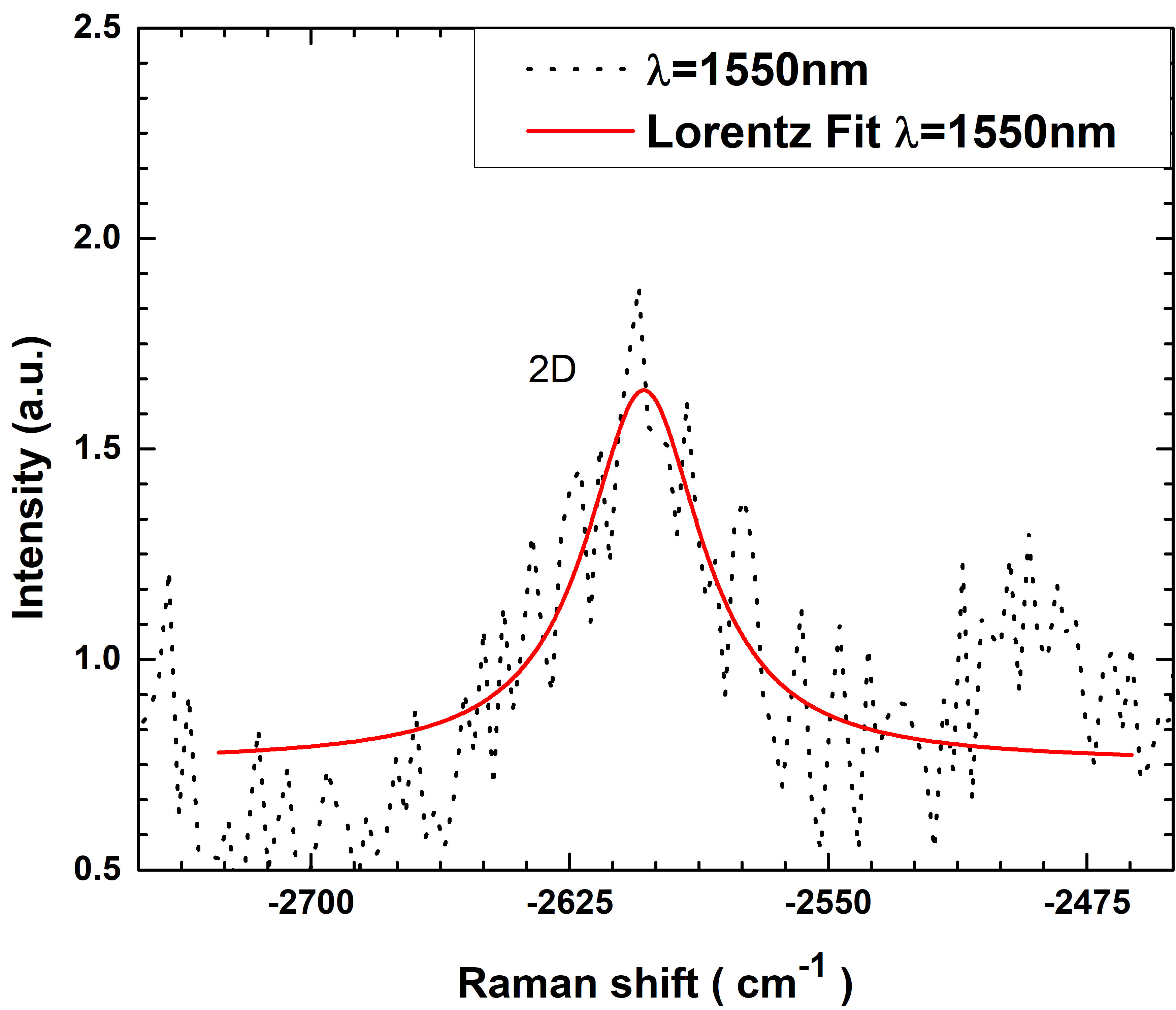}
    \caption{ Detection of signal with fitted Lorentzian peak at $-2610$ cm$^{-1}$ under $1550$ nm excitation, after background removal.}
    \label{fig:2D peak}
\end{figure}

In the MRR device used in setup 2, graphene needs to be deposited as a solution into the etched hole, so that it reaches the surface of the ring. A similar laser held at $1550$ nm (Amonics, ATL-C-16-B-FA) was used as excitation source by being fed into the device through the ``in'' port. Laser power was measured to be around $20$ mW.  A signal at $-4220$ cm$^{-1}$ is observed [Figure 3 (b), main text] after background subtraction, and is discussed further in the main text. The artifacts from the amplifier do not affect the peaks in this range.\\
Using setup 2, we also probed range of other wavelengths to look for Raman signals with $1600$ nm excitation, but due to the limitation of the in-house modified Raman system, the graphene Raman shifts from further away were not detectable. We see strong parasitic signal from the the amplifier and optical fiber cables used to couple input laser and the spectrometer probe during all measurements as discussed in above section.  Still, after subtracting the background obtained by measuring empty MRR spectra and removing isolated noise-peaks, a potential signal is seen around $-2610$ cm$^{-1}$ as shown here in Figure S3. This is a candidate for 2D Raman shift of graphene under $1550$ nm excitation~\cite{Ferrari2006sup,ferrari2013sup,Ferrante2018sup}. High noise level in the resulting signal can be attributed to combination of factors; firstly, the etching of a hole on the device likely causes variations in refractive index around the ring which creates disturbances to the scattering processes happening on the ring. Secondly, as seen in Figure S2 (a), the background intensity is very high in this device at the location of potential 2D peak. Hence, any potential signal is highly suppressed but is still recoverable in this case.

\end{document}